\documentclass{aastex6}
\usepackage{graphicx}
\begin{document}

\title{Constraints on the Cosmic-Ray Ionization Rate in the $z\sim2.3$ Lensed Galaxies SMM~J2135$-$0102 and SDP~17b from Observations of OH$^+$ and H$_2$O$^+$}

\author{Nick Indriolo}
\affil{Space Telescope Science Institute, Baltimore, MD 21218, USA}
\author{E.~A.~Bergin}
\affil{University of Michigan, Ann Arbor, MI 48109, USA}
\author{E.~Falgarone}
\author{B. Godard}
\affil{LERMA, Observatoire de Paris, Ecole Normale Sup\'{e}rieure, PSL Research University, CNRS, UMR8112, F-75014 Paris, France}
\author{M.~A.~Zwaan}
\affil{European Southern Observatory, Karl-Schwarzchild-Strasse 2, 85748 Garching, Germany}
\author{D.~A.~Neufeld}
\affil{Department of Physics and Astronomy, Johns Hopkins University, Baltimore, MD 21218, USA}
\author{M.~G.~Wolfire}
\affil{Department of Astronomy, University of Maryland, College Park, MD 20742, USA}

\begin{abstract}
Cosmic rays are predominantly accelerated in shocks associated with star formation such as supernova remnants and stellar wind bubbles, so the cosmic-ray flux and thus cosmic-ray ionization rate, $\zeta_{\rm H}$, should correlate with the star-formation rate in a galaxy. Submillimeter bright galaxies (SMGs) are some of the most prolific star forming galaxies in the Universe, and gravitationally lensed SMGs provide bright continuum sources suitable for absorption line studies. Abundances of OH$^+$ and H$_2$O$^+$ are useful for inferring $\zeta_{\rm H}$ when combined with chemical models, and have been used for this purpose within the Milky Way. At redshifts $z\gtrsim2$ transitions out of the ground rotational states of OH$^+$ and H$_2$O$^+$ are observable with ALMA, and we present observations of both molecules in absorption toward the lensed SMGs SMM~J2135$-$0102 and SDP~17b. These detections enable an exploration of $\zeta_{\rm H}$ in galaxies with extreme star formation and high supernova rates, both of which should significantly enhance cosmic-ray production. The observed OH$^+$ and H$_2$O$^+$ absorption is thought to arise in massive, extended halos of cool, diffuse gas that surround these galaxies. Using a chemical model designed to focus on the reaction network important to both species, we infer cosmic-ray ionization rates of $\zeta_{\rm H}\sim10^{-16}$--$10^{-14}$~s$^{-1}$ in these extended gaseous halos. As our estimates come from gas that is far away from the sites of cosmic-ray acceleration, they imply that cosmic-ray ionization rates in the compact regions where star formation occurs in these galaxies are orders of magnitude higher.
\end{abstract}

\section{Introduction} \label{sec_intro}

The star formation rate (SFR) density has varied throughout the history of the Universe, peaking around a redshift of $z\approx2$ \citep{madau2014}. Submillimeter-bright galaxies (SMGs) observed during this epoch of peak star formation outshine even ultra-luminous infrared galaxies (ULIRGs) and starburst galaxies, and it has been speculated that they are the progenitors of the massive elliptical galaxies seen in the local Universe \citep[e.g.,][and references therein]{simpson2014}. Characterizing the physical conditions in SMGs is thus important for better understanding star formation in and evolution of massive galaxies.

Star formation is regulated by the molecular gas content in a galaxy \citep{wong2002}, and in turn, the newly formed stars influence the physical and chemical conditions in the interstellar medium (ISM) through various feedback mechanisms. Many atomic and molecular species are useful in constraining properties of the ISM such as density, temperature, and radiation field since the relative population in different quantum states is controlled by radiative and collisional (de)-excitation. As a result, observations of emission and absorption from species such as CO, H$_2$O, C, and CH$^+$ can and have been used to infer properties of SMGs \citep[e.g.,][]{swinbank2011,falgarone2017,yang2017}.

One property of SMGs that has received minimal study thus far is the cosmic-ray ionization rate, and the underlying cosmic-ray flux. Because photons capable of ionizing hydrogen ($E>13.6$~eV) do not penetrate very far into molecular clouds, cosmic rays represent the dominant ionization mechanism in cloud interiors. Interstellar gas phase chemistry is primarily driven by fast ion-molecule reactions, so this ionization source is vital to increasing chemical complexity. Cosmic rays are accelerated in shocks associated with massive stars---most notably in supernova remnants, but also in the termination shocks of stellar wind bubbles---so it follows that the cosmic-ray flux should be correlated with the star formation rate. Indeed, estimates of the cosmic-ray ionization rate increase from the Galactic disk \citep[$10^{-16}$~s$^{-1}$;][]{indriolo2012,indriolo2015oxy} to the Galactic center \citep[$10^{-15}$--$10^{-13}$~s$^{-1}$; e.g.,][]{yusefzadeh2013,goto2014,lepetit2016} to nearby star-forming galaxies \citep[$10^{-13}$--$10^{-12}$~s$^{-1}$;][]{gonzalez-alfonso2013,gonzalez-alfonso2018arxiv}. Our goal is to infer the cosmic-ray ionization rate in SMGs at $z\sim2.3$ to better understand the impact of prolific star formation on the gas reservoirs in these massive galaxies, and to trace the relation between star formation and ionization rate in the early Universe.

Multiple studies of OH$^+$ and H$_2$O$^+$ in the Milky Way have demonstrated that these oxygen-bearing ions are useful in constraining the cosmic-ray ionization rate of atomic hydrogen, $\zeta_{\rm H}$ \citep{gerin2010,neufeld2010,porras2014,indriolo2015oxy,zhao2015}. This is because the formation, and thus abundances, of these species is closely linked to the ionization of atomic hydrogen through the following chemical reactions: 
\begin{equation}
{\rm H} + {\rm CR} \rightarrow {\rm H}^+ + e^- + {\rm CR'},
\label{reac_CR_H}
\end{equation} 
\begin{equation}
{\rm H}^+ + {\rm O} + \Delta E \longleftrightarrow {\rm O}^+ + {\rm H},
\label{reac_Hp_Op}
\end{equation}
\begin{equation}
\mathrm{O^+ + H_2 \rightarrow OH^+ + H},
\label{reac_Op_H2}
\end{equation}
\begin{equation}
\mathrm{OH^{+} + H_{2} \rightarrow H_{2}O^{+} + H}.
\label{reac_OH+_H2}
\end{equation}
The reaction network surrounding these species is, of course, more complicated than these four reactions \citep[see, e.g.,][]{hollenbach2012,indriolo2015oxy}, and the reactions shown above are only intended to demonstrate how the production of OH$^+$ and H$_2$O$^+$ is a few steps removed from the ionization of H. Given our success with inferring cosmic-ray ionization rates from OH$^+$ and H$_2$O$^+$ in the Galaxy, we have targeted transitions from the ground rotational states of both species in a small sample of SMGs at $z\sim2.3$ with the same goal in mind. Throughout this paper we will refer to the cosmic-ray ionization rate of atomic hydrogen, $\zeta_{\rm H}$, which accounts for ionization by both cosmic rays and energetic secondary electrons produced via ionization. Estimates of the primary ionization rate of atomic hydrogen, $\zeta_p$, which does not include ionization by secondary electrons, and the ionization rate of molecular hydrogen, $\zeta_2$, can be made via the relations $\zeta_{\rm H}=1.5\zeta_p$ and $\zeta_2=2.3\zeta_p$ \citep{glassgold1973,glassgold1974}.

\section{Observations}

\subsection{Target Selection}
Over the past decade, far-infrared photometric surveys (e.g., H-ATLAS) have identified a sample of SMGs that are particularly bright due to magnification from gravitational lensing \citep{negrello2017}. Spectroscopic redshifts have been determined for several such objects \citep[e.g.,][]{harris2012,lupu2012}, enabling targeted observations of transitions from various atomic and molecular species. Some of the most commonly observed species are CO, H$_2$O, and C, all of which are seen in emission in the star-forming regions of SMGs \citep[e.g.,][]{omont2013,yang2016,yang2017}. Recent observations have detected CH$^+$ in both emission and absorption in lensed SMGs \citep{falgarone2017}. CH$^+$ absorption generally traces cool ($T\lesssim 100$~K), diffuse ($n\lesssim 100$~cm$^{-3}$) gas where hydrogen is not fully molecular, a different environment than that giving rise to the CO and H$_2$O emission. These are, however, conditions that favor the formation of OH$^+$ and H$_2$O$^+$, and principal component analysis of Galactic sight lines has demonstrated that CH$^+$, OH$^+$, and H$_2$O$^+$ are all most commonly observed in low density gas where hydrogen is predominantly in atomic form \citep{gerin2016}. To study the cosmic-ray ionization rate using oxygen-bearing ions at $z\sim2.3$ we selected galaxies from \citet{falgarone2017} that show CH$^+$ absorption as potential targets. In this paper we focus on the observations of SMM~J2135$-$0102 and SDP~17b, and list some general properties of these galaxies in Table \ref{tbl_targets}. 

At $z\sim2.3$ the ground-state transitions of OH$^+$ and H$_2$O$^+$ that have rest-frame frequencies near 1~THz (see Table \ref{tbl_transitions}) are shifted down to  frequencies covered by band 7 of ALMA. There are two sets of hyperfine transitions that probe the ground state of H$_2$O$^+$ near 1115~GHz and 1139~GHz, with the former separated from the ground state transition of water at 1113.343~GHz by only about 500~km~s$^{-1}$. Molecular emission lines in these galaxies typically have widths of a few hundred km~s$^{-1}$, so we chose to target the 1139~GHz complex of H$_2$O$^+$ to avoid confusion caused by potential blending of the 1115~GHz complex with water vapor. Three sets of hyperfine transitions probe the ground state of OH$^+$ at 909~GHz, 971~GHz, and 1033~GHz. Observations covering the 1033~GHz transition in both target galaxies are publicly available in the ALMA archive.

\subsection{Execution}

Two of the observations presented here were made as part of program ADS/JAO.ALMA\#2016.1.00285.S: H$_2$O$^+$ at 1139~GHz in SMM~J2135$-$0102 and SDP~17b. Three basebands, each with a single spectral window, were placed in the same sideband to provide continuous spectral coverage across the widest possible frequency range (about 3000~km~s$^{-1}$), with the middle window centered on the strongest hyperfine component of the targeted transition. A frequency resolution of 31.25~MHz provides velocity resolution of about 30~km~s$^{-1}$. SDP~17b was observed for 1481~s on 2016 Nov 26 targeting H$_2$O$^+$, with J0854+2006 as the bandpass calibrator, J0909+0121 as the phase calibrator, and J1058+0133 as the flux calibrator. SMM~J2135$-$0102 was observed for 1451~s on 2017 Apr 14 targeting H$_2$O$^+$, with J2134$-$0153 as the phase calibrator, J2232+1143 as the bandpass calibrator, and Titan as the flux calibrator.

The observations targeting OH$^+$ at 1033~GHz toward SMM~J2135$-$0102 and SDP~17b were made as parts of programs ADS/JAO.ALMA\#2012.1.00175.S and ADS/JAO.ALMA\#2013.1.00569.S, respectively. Both placed 2 basebands with a single spectral window each in the upper sideband to provide near-continuous coverage over 2500-3000~km~s$^{-1}$ at 31.25~MHz spectral resolution. SMM~J2135$-$0102 was observed for 3568~s on 2014 Aug 19 targeting OH$^+$ with J2134$-$0153 as the phase calibrator and J2148+0657 as the bandpass, amplitude, and flux calibrator. SDP~17b was observed for 1965~s on 2015 Jun 15 targeting OH$^+$, with J0909+0121 as the phase calibrator and J1058+0133 as the bandpass, amplitude, and flux calibrator.

\section{Data Reduction}

\begin{figure}
\epsscale{1.0}
\plotone{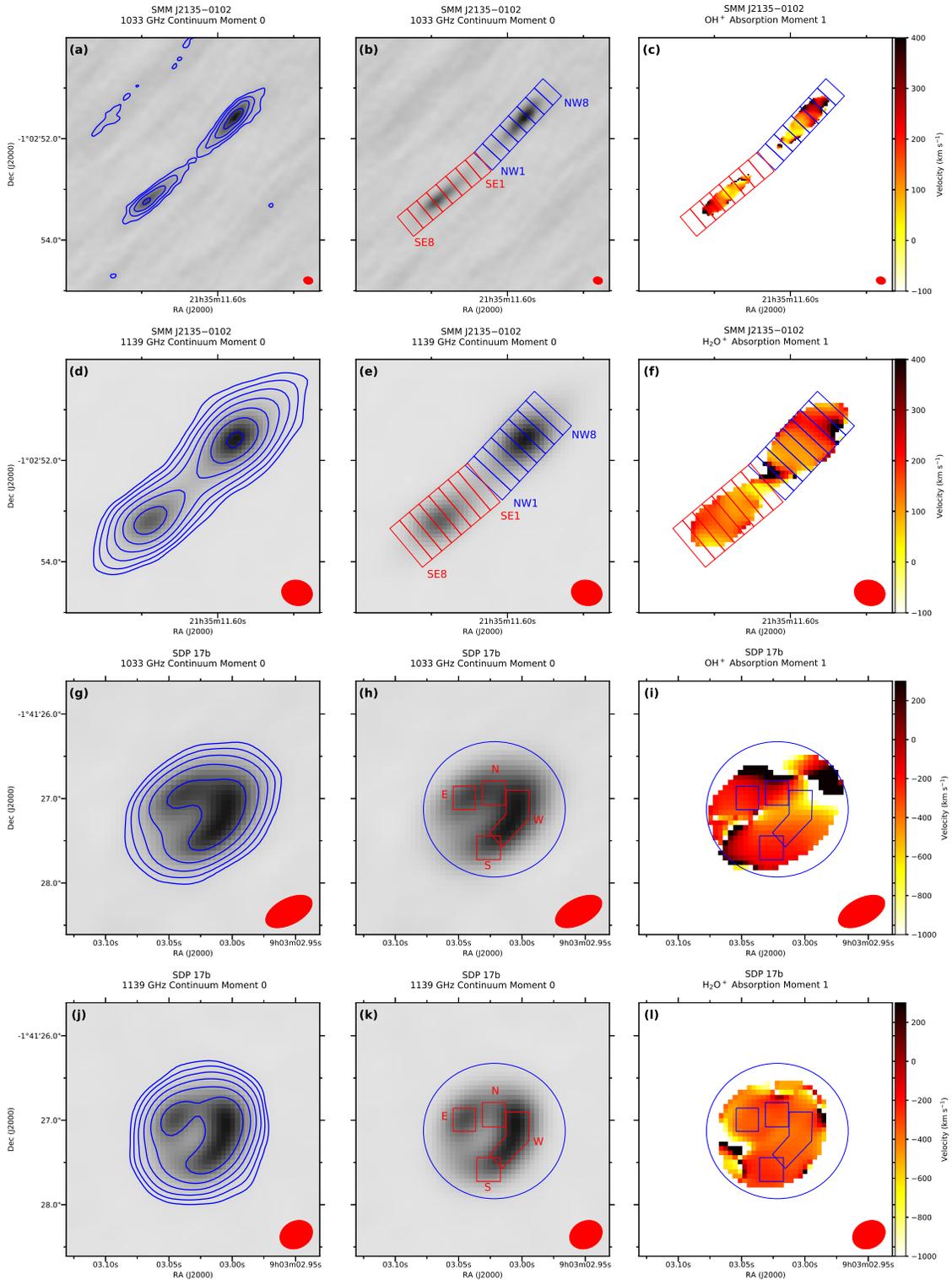}
\caption{Moment maps of the two target galaxies at frequencies near the OH$^+$ and H$_2$O$^+$ transitions. The left and middle columns show the integrated continuum intensity as grayscale images. Synthesized beams are shown in the lower right corners of each panel. In panels a, d, g, and j contours show $5\sigma\times2^n$, where $\sigma$ is the root-mean-squared noise level, $n=0, 1, 2, ..., m$, and $m$ is the last such value below the peak flux. Panels b, e, h, and k show the regions from which spectra were extracted. In SDP~17b the blue ellipse marks the region used to extract the spectrum over the entire source. For SMM~J2135$-$0102, the full NW and SE spectra are from the regions comprising all blue and red rectangles, respectively. Panels c, f, i, and l show first moment maps of the OH$^+$ and H$_2$O$^+$ absorption (i.e., continuum-subtracted, absorption-weighted velocity) in each source.}
\label{fig_continuum_images}
\end{figure}

All data calibration was performed by NRAO staff using standard procedures to produce calibrated measurement sets. Further processing starting from those products was done with CASA version 5.0.0. One round of phase-only self-calibration using absorption-free portions of the continuum was applied to all observations (except for that covering OH$^+$ in SMM~J2135$-$0102) in order to reduce noise levels. Measurement sets were imaged via the \texttt{clean} task in CASA utilizing the Briggs weighting scheme with a robust parameter of 0.5. These cleaned, calibrated data sets were used to produce position-position-velocity data cubes centered at the strongest hyperfine component of the targeted transition given the redshift of each source. Moment 0 maps of the continuum emission from each observation are shown in Figure \ref{fig_continuum_images}. The spatial distribution of continuum emission at rest-frame frequencies near 1033~GHz and 1139~GHz resembles that observed at 835~GHz \citep{falgarone2017}. For SMM~J2135$-$0102, however, we do not see the clumpy structure previously reported at similar frequencies with comparable angular resolution \citep[870~$\mu$m observed; 1145~GHz rest-frame;][]{swinbank2010}.

Spectra of the target galaxies were extracted from the data cubes using the Spectral Profile Tool within CASA Viewer. Many previous observations of our target SMGs do not spatially resolve the continuum emission from these galaxies, so we present and analyze spectra extracted over the full extent of the objects for comparison with those studies. Given the high angular resolution of our ALMA observations, we have also extracted spectra from various sub-regions within the target galaxies as marked in Figure \ref{fig_continuum_images}.  All of the resulting spectra are presented in Figures \ref{fig_eyelash_ohp_spectra_v2} through \ref{fig_sdp17b_spectra}. Velocity scales use the radio definition and are referenced to the strongest hyperfine component of each transition (see Table \ref{tbl_transitions}) at the redshifts listed in Table \ref{tbl_targets}. Top panels show spectra from regions that encompass the full spatial extent of continuum emission for each source, while the other panels show spectra extracted from the various sub-regions. Note that while the horizontal (velocity) axis is constant for all panels in each figure, the vertical (flux density) axis varies between panels.

\section{Resulting Spectra}

\subsection{SMM~J2135$-$0102}

\begin{figure}
\epsscale{1.0}
\plotone{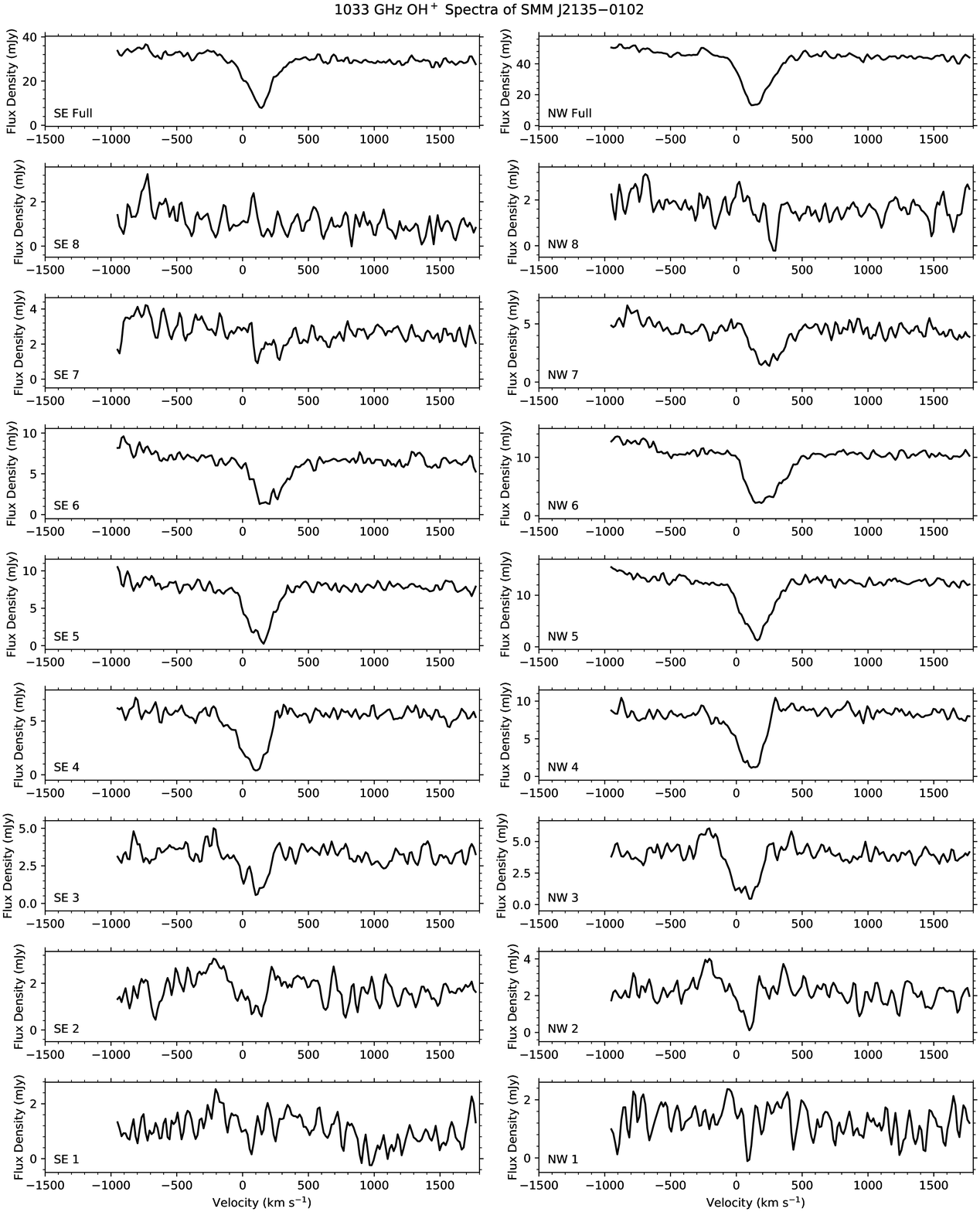}
\caption{Spectra of SMM~J2135$-$0102 covering the 1033~GHz transition of OH$^+$ in the sub-regions marked in Figure \ref{fig_continuum_images}. Panels on the left-hand side show spectra from the sub-regions of the SE image of the galaxy, while panels on the right-hand side show spectra from the sub-regions of the NW image of the galaxy. The top panels (labeled Full) show spectra extracted from regions defined by the outer boundaries of all of the sub-regions.}
\label{fig_eyelash_ohp_spectra_v2}
\end{figure}

\begin{figure}
\epsscale{1.0}
\plotone{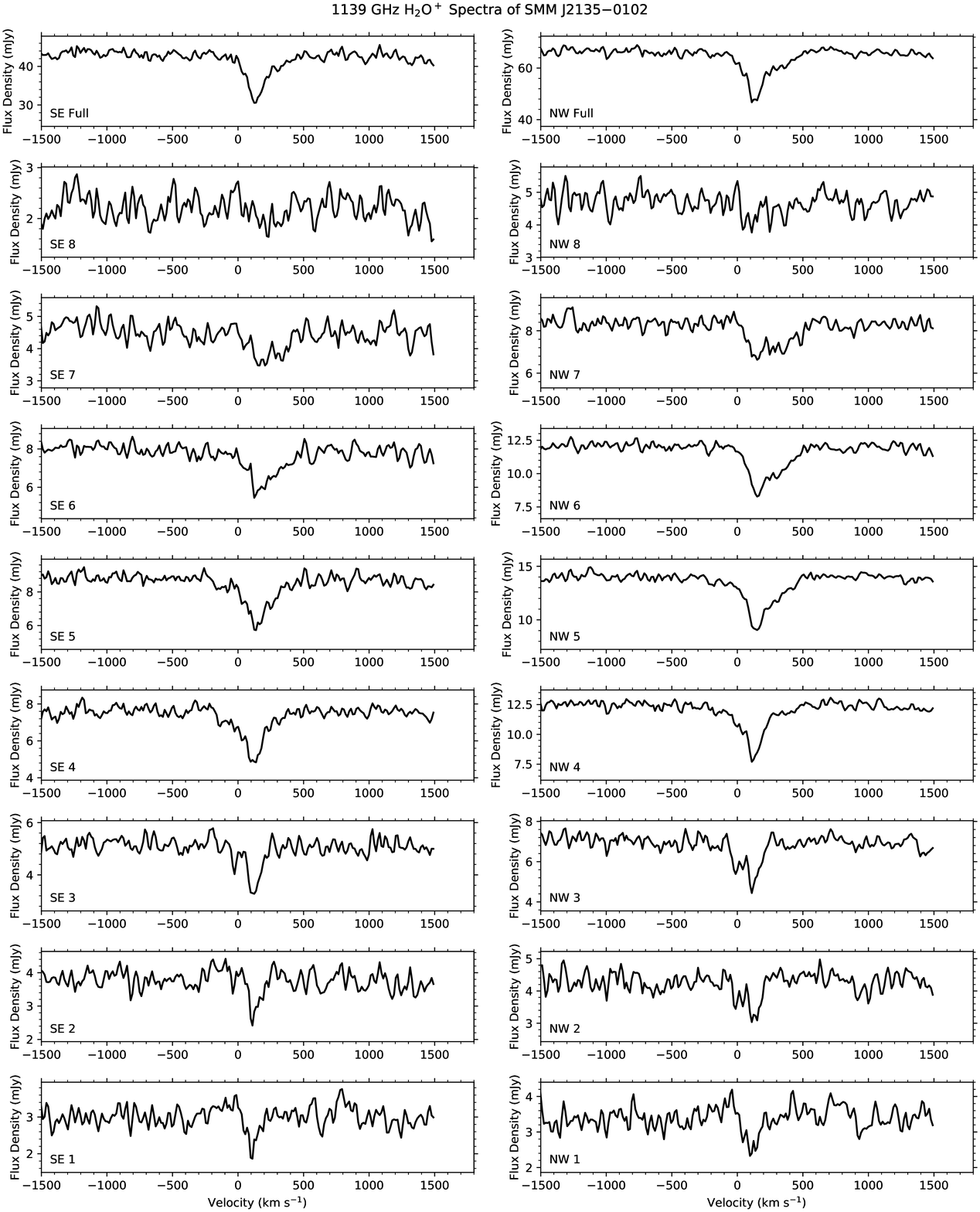}
\caption{Same as Figure \ref{fig_eyelash_ohp_spectra_v2}, but for the 1139~GHz transition of H$_2$O$^+$.}
\label{fig_eyelash_h2op_spectra_v2}
\end{figure}

Figures \ref{fig_eyelash_ohp_spectra_v2} and \ref{fig_eyelash_h2op_spectra_v2} show OH$^+$ and H$_2$O$^+$ spectra toward SMM~J2135$-$0102. The NW and SE images of the galaxy are reflected about the $z=2.3$ critical curve \citep{swinbank2010,swinbank2011}, and the 8 sub-regions sample roughly the same location along the major axis in the two images (see Figure \ref{fig_continuum_images}).  While the continuum level flux density is about 1.5 times higher in the NW image than the SE image, the OH$^+$ and H$_2$O$^+$ absorption line profiles for the mirrored regions agree with each other within the noise levels. It is clear that the absorption profiles of OH$^+$ and H$_2$O$^+$ change with position along the major axis of the galaxy, shifting to larger velocities with increasing distance from the critical curve. This is visible in the first moment maps of OH$^+$ and H$_2$O$^+$ absorption shown in Figure \ref{fig_continuum_images} panels (c) and (f), respectively, as a velocity gradient along the major axis of the galaxy.  Position-velocity maps of CH$^+$ absorption \citep{falgarone2017} and of CO $J=6$--5 and 1--0 emission \citep{swinbank2011} show a similar behavior. Together, these observed velocity gradients indicate large-scale rotation of both the warm gas and dust where CO 6--5 and rest-frame 1~THz continuum emission arises, as well as the cool foreground gas giving rise to OH$^+$, H$_2$O$^+$, and CH$^+$ absorption.  Correlation between the strength of continuum emission and foreground absorption further suggests a relationship between the two components.

The broad component of CH$^+$ emission seen in SMM~J2135$-$0102 is not definitively detected in OH$^+$. There is a hint of broad OH$^+$ emission in some of the spectra (see, e.g., NW2, NW3, and NW4 in Figure \ref{fig_eyelash_ohp_spectra_v2}) between about $-$500~km~s$^{-1}$ and 1000~km~s$^{-1}$, but it is difficult to determine if this feature is real without wider frequency coverage. It is also possible that the small synthesized beam used for the OH$^+$ observations (0\farcs18$\times$0\farcs14) resolves out the emission seen in CH$^+$ observations. There does, however, appear to be a narrow (FWHM$\sim$100~km~s$^{-1}$) component of OH$^+$ emission centered at about $-$250~km~s$^{-1}$ toward some of the regions closer to the $z=2.3$ critical curve (e.g., NW2, NW3, SE2, SE3). The final emission signature seen in the OH$^+$ spectra of SMM~J2135$-$0102 is at the extreme blue end ($-$700~km~s$^{-1}$ to $-$900~km~s$^{-1}$) toward regions NW5, NW6, NW7, SE5, and SE6. This emission is due to the CO $J=9$--8 transition at 1036.9124~GHz and was previously reported by \citet{danielson2011}. There is no indication of H$_2$O$^+$ emission in SMM~J2135$-$0102.

\subsection{SDP~17b}

Spectra of OH$^+$ and H$_2$O$^+$ toward SDP~17b are shown in the left and right portions of Figure \ref{fig_sdp17b_spectra}, respectively. The gap between about 800 and 1100~km~s$^{-1}$ in the OH$^+$ spectra is due to non-continuous frequency coverage between the adjacent spectral windows. The OH$^+$ line profile in the spectrum covering the entire source resembles that of CH$^+$ \citep{falgarone2017}, with strong absorption from about $-$800 to 200~km~s$^{-1}$, and what appears to be part of a broad emission component from about 200 to 800~km~s$^{-1}$. A quantitative analysis of this emission component is severely impacted by the aforementioned gap in coverage and the lack of data beyond $-$800~km~s$^{-1}$, so we do not attempt such an analysis. The OH$^+$ absorption profile changes between the different sub-regions, most notably reaching 75\% absorption in the South region---a region that accounts for less than 10\% of the total continuum flux density---while only reaching about 50\% absorption elsewhere. It is unlikely that there is a contribution from CO 9--8 in these spectra as observations covering both the CO 7--6 and 8--7 transitions resulted in non-detections \citep{yang2017}. The H$_2$O$^+$ absorption profile in SDP~17b is both weaker and narrower than that of OH$^+$, although its full extent in velocity is difficult to determine given the relatively low signal-to-noise (S/N) level of the continuum. Like OH$^+$, H$_2$O$^+$ absorption is strongest in the South region. First moment maps of OH$^+$ and H$_2$O$^+$ absorption (Figure \ref{fig_continuum_images} panels (i) and (l), respectively) do not show any rotation signature, and the extreme velocities at the edges of these maps are the result of low S/N in the spectra extracted from those pixels.

\section{Analysis}
 
\subsection{Column Densities} \label{section_column}

\begin{figure}
\epsscale{1.0}
\plotone{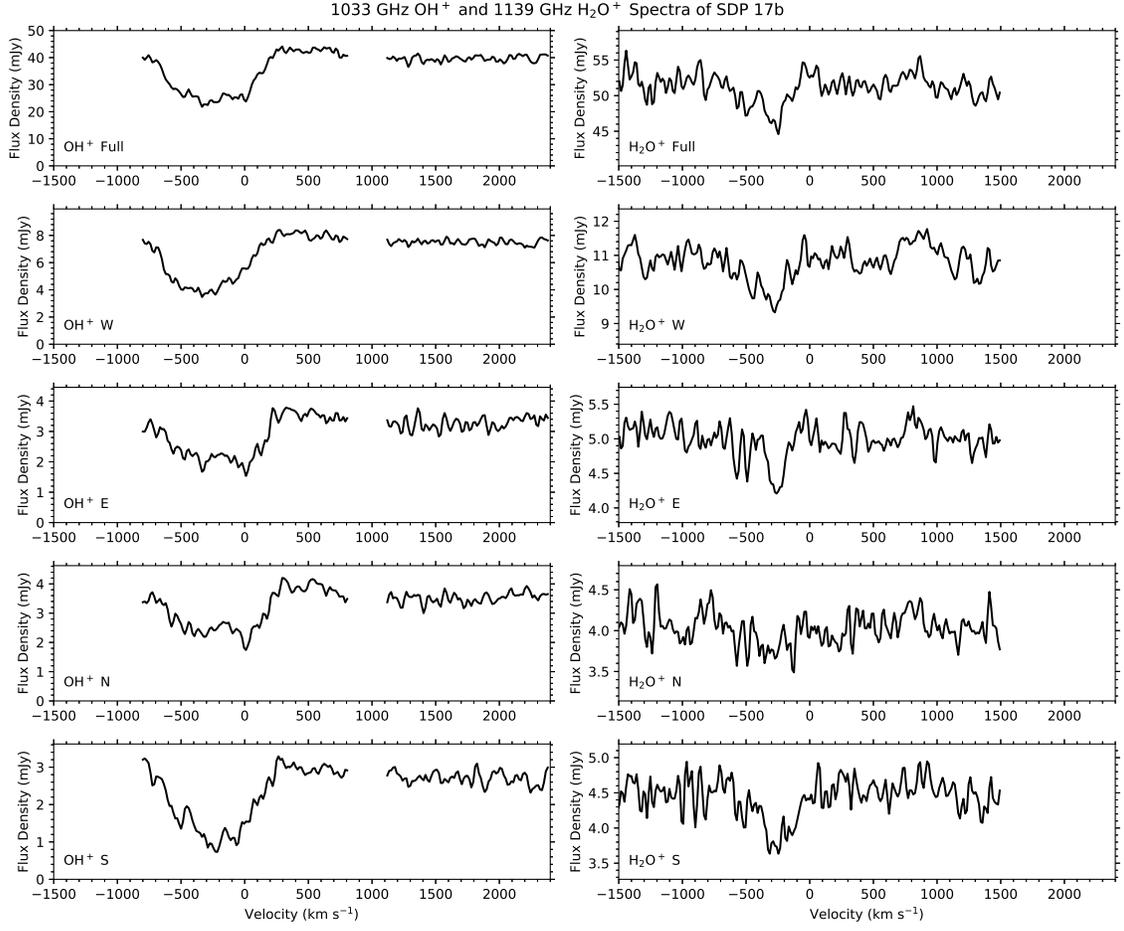}
\caption{Spectra of SDP 17b covering the 1033~GHz transition of OH$^+$ (left) and the 1139~GHz transition of H$_2$O$^+$ (right). The top panels show spectra extracted from the region that encompasses all of the continuum emission, while the others show spectra from the various sub-regions defined in Figure \ref{fig_continuum_images}. The gap near 1000~km~s$^{-1}$ in the OH$^+$ spectra is due to non-continuous frequency coverage between two spectral windows.}
\label{fig_sdp17b_spectra}
\end{figure}

In order to determine column densities of OH$^+$ and H$_2$O$^+$, the spectra presented in Figures \ref{fig_eyelash_ohp_spectra_v2} through \ref{fig_sdp17b_spectra} are first normalized to a scale where unity represents complete transmission and zero is complete absorption. This is done by dividing the spectra by a moving average that was interpolated across velocities showing absorption or narrow emission. Normalized transmission spectra are then converted to optical depth spectra using the standard relation $\tau=-\ln({\rm transmission})$. Velocity ranges were defined to fully encompass absorption features and are reported in Table \ref{tbl_measure} along with the integrated optical depths ($\int\tau dv$) measured over those velocity intervals. Uncertainties in the integrated optical depths are estimated from the standard deviation of the continuum level of transmission spectra. The standard deviation of the normalized continuum is determined over a line-free region, and this is assumed to be the error in each channel. This error is divided by the transmission level to compute the $1\sigma$ uncertainty in optical depth in each channel, and the uncertainty in the integrated optical depth is taken to be the product of $\Delta v$ (the channel width in km~s$^{-1}$) and the square root of the sum of variances in optical depth for each channel that was included in the integration.

Column densities in the rotational states from which the observed transitions arise---in these cases the ground rotational states of OH$^+$ and H$_2$O$^+$---are calculated from integrated optical depths as
\begin{equation}
N_{l}=\frac{8\pi}{c^3}\sum_S g_l\sum_{T}A_{ul}^{-1}g_u^{-1}\nu^{3} \int\tau dv,
\label{eq_column}
\end{equation}
where $c$ is the speed of light, $g_l$ and $g_u$ are the lower and upper state statistical weights, respectively, $A_{ul}$ is the spontaneous emission coefficient, $\nu$ is the transition frequency, the first sum is over $g_l$ in all of the sub-states that make up the ground rotational state, and the second sum is over $A_{ul}^{-1}g_u^{-1}\nu^{3}$ for all of the transitions that are contributing to absorption (see Table \ref{tbl_transitions} for transition properties). As the energy difference between the sub-states is negligible, their relative populations can be approximated solely by their statistical weights. This equation also makes the simplifying assumption that the absorption from all hyperfine components is at the same frequency. Given the width of the absorption features and the instrumental spectral resolution, ignoring the of-order 30~km~s$^{-1}$ shift between the different hyperfine transitions is a reasonable approximation.  To compute column density in units of cm$^{-2}$ when $\int\tau dv$ has units of km~s$^{-1}$, the constants outside of the integral in equation (\ref{eq_column}) are
$4.87\times10^{12}$ and $4.18\times10^{12}$ for the 1033~GHz and 1139~GHz transitions, respectively.

Determining total column densities for a species requires accounting for the population in excited states in addition to the ground state. OH$^+$ and H$_2$O$^+$ absorption primarily arises in low density ($n_{\rm H}<100$~cm$^{-3}$) gas \citep{indriolo2015oxy} where collisional excitation is slow compared to spontaneous emission. The cosmic microwave background (CMB) at $z=2.3$ has a blackbody temperature of $T_{\rm CMB}\approx9$~K, and so does not significantly populate the first rotationally excited state of OH$^+$ ($\sim45$~K). As such, the column density in the ground rotational state is a good approximation of the total OH$^+$ column density. The case for H$_2$O$^+$ is more complicated as the two identical protons lead to both {\it ortho} and {\it para} nuclear spin modifications that do not communicate via radiative transitions or non-reactive collisions, meaning there are effectively two separate ground states. The first excited state of $o$-H$_2$O$^+$ is about 55~K above the ground {\it ortho} state (the state probed by the 1139~GHz transition) such that it is not significantly populated by CMB radiation. The {\it ortho}-to-{\it para} ratio expected based solely on statistical weights is 3:1, and we assume this value when computing the total H$_2$O$^+$ column density from our observations of {\it ortho}-H$_2$O$^+$. Total column densities of OH$^+$ and H$_2$O$^+$ determined from all of our spectra are reported in Table \ref{tbl_measure}, with uncertainties computed from the uncertainties in integrated optical depth discussed above. However, it is possible that we have underestimated column densities if saturated absorption is being masked by other effects. These might include emission from the upper state in the observed transitions filling in part of the line, narrow (in velocity), spectrally unresolved absorption components, or non-uniform and/or incomplete screening of the background continuum by the absorbing gas. Indeed, we likely see the last effect toward SDP~17b (compare the absorption depth of OH$^+$ in the Full and South spectra in the top-left and bottom-left panels of Figure \ref{fig_sdp17b_spectra}, respectively), and it may operate on smaller spatial scales within the defined sub-regions as well. The column densities reported in Table \ref{tbl_measure} then, are really the averages of all column densities in front of infinitesimally small solid angles ($d\Omega$) of the background continuum region, each weighted by the continuum flux within $d\Omega$.\footnote{This is universally true for column densities determined from absorption lines, but it is worth explicitly stating here as we know that the continuum emission varies within our defined sub-regions.} Still, it is possible that these are only lower limits if saturated absorption is masked by the above effects.

\subsection{Ionization Rates} \label{section_ionizationrates}

The formation of OH$^+$ and H$_2$O$^+$ begins with the ionization of H, and the abundances of our target species can be used in constraining the ionization rate. While analytical expressions derived from steady-state chemistry have frequently been used for this purpose \citep[e.g.,][]{indriolo2015oxy}, detailed chemical models have shown that those expressions are only valid over a limited range of physical conditions \citep{neufeld2017}. In particular, the expressions are invalid at high ionization rates ($\zeta_{\rm H}\gtrsim10^{-15}$~s$^{-1}$) where the ionization of H becomes the primary source of free electrons, and the OH$^+$ abundance no longer scales linearly with $\zeta_{\rm H}$ \citep{indriolo2015oxy}. Additionally, those expressions require estimates of the atomic hydrogen column density ($N({\rm H})$), hydrogen nucleon density ($n_{\rm H}\equiv n({\rm H})+2n({\rm H}_2)$), electron fraction ($x_e\equiv n_e/n_{\rm H}$), and OH$^+$ production efficiency ($\epsilon$), all of which are either poorly constrained or unconstrained in our target galaxies. We thus choose to employ the chemical models presented in \citet{neufeld2017}, removing the need to adopt unconstrained values of $x_e$ and $\epsilon$.

The analysis of \citet{neufeld2017}, first presented by \citet{hollenbach2012}, utilizes a chemical network designed specifically to focus on processes important to oxygen chemistry to predict the abundances of OH$^+$ and H$_2$O$^+$ in a model interstellar cloud. We provide a brief description of the chemical model here, and for more details the interested reader is referred to \citet{hollenbach2012} and \citet{neufeld2017}, and references therein. The model cloud is a plane parallel slab with uniform gas density that is illuminated on both sides by a UV radiation field. Certain parameters, including the cosmic-ray ionization rate, metallicity ($Z$), hydrogen nucleon density, total visual extinction through the cloud ($A_V$), and ultraviolet radiation field \citep[$\chi_{\rm UV}$, in multiples of the mean interstellar radiation field from][]{draine1978}, are specified as input. Equilibrium temperature and steady-state chemical abundances are computed as a function of depth into the cloud, and integration through the cloud provides predicted column densities for a given set of model input parameters. By running the chemical model using several different combinations of input parameters we can investigate clouds with a range of physical conditions.

The range of input parameters that we chose to explore is as follows. Metallicities in submillimeter galaxies at $z\sim2$ have been estimated to be roughly solar \citep{bothwell2016}, so we use the chemical models with $Z=Z_\sun$ in our analysis. CH$^+$ absorption in the Galactic disk primarily arises in gas with 30~cm$^{-3}\leq n_{\rm H}\leq 50$~cm$^{-3}$, and the same is likely true of the absorption presented by \citet{falgarone2017} so we adopt models with $n_{\rm H}=50$~cm$^{-3}$. The UV radiation fields in SMGs are estimated to be 10$^2$--10$^4$ times stronger than in the Milky Way \citep{wardlow2017} due to prolific star formation. However, those estimates come from dense, molecular gas that is in close proximity to the regions where stars are forming. \citet{falgarone2017} concluded that the CH$^+$ absorption they observe arises from massive, low-density gas halos that extend to of order 10~kpc around these compact ($\sim1$~kpc diameter) galaxies. The intensity of the radiation field will decrease with increasing distance from the sites of star formation, and nearly all of the UV flux will be attenuated outside of the dusty starburst regions. Because it is not clear exactly where the gas giving rise to OH$^+$ and H$_2$O$^+$ absorption is located in these galaxies, we run the suite of chemical models using $\chi_{\rm UV}$=0.1, 1, 10, 100, and 1000 to investigate a range of potential UV radiation fields. Finally, we sample the parameter space bounded by $0.0003$~mag $\leq A_V \leq 8$~mag and $9\times10^{-19}$~s$^{-1}\leq \zeta_{\rm H}\leq 9\times10^{-15}$~s$^{-1}$ \citep[specific values are listed in][]{neufeld2017} to explore a range of cloud thicknesses and cosmic-ray ionization rates. 

\begin{figure}
\epsscale{1.0}
\plotone{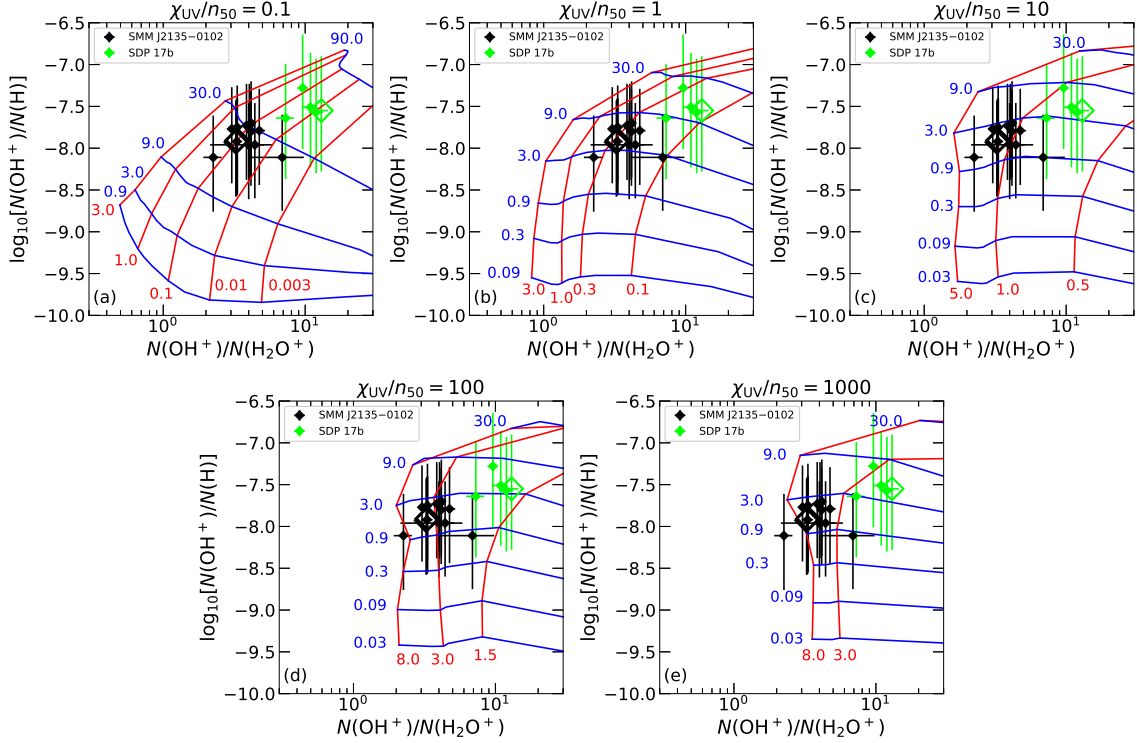}
\caption{Contours of constant cosmic-ray ionization rate of atomic hydrogen in units of $10^{-16}$~s$^{-1}$ ($\zeta_{\rm H}/n_{50}$; blue) and of total visual extinction through the model cloud  in units of magnitudes ($A_V$; red) are plotted in the parameter space $\log_{10}[N({\rm OH}^+)/N({\rm H})]$ vs. $N({\rm OH}^+)/N({\rm H_2O}^+)$ as determined from a suite of chemical models presented in \citet{neufeld2017}. Each panel shows the model results assuming different strengths of the UV radiation field ($\chi_{\rm UV}/n_{50}$) impinging on the cloud with plane-parallel slab geometry. The density in all model runs is $n_{\rm H}=50$~cm$^{-3}$. Black (SMM~J2135$-$0102) and green (SDP~17b) data points are from $N({\rm OH}^+)$ and $N({\rm H_2O^+})$ determined from our ALMA observations and from $N({\rm H})$, estimated from $N({\rm CH^+})$, $X({\rm CH}^+)$, and $f_{\rm H_2}$ as discussed in Section \ref{section_ionizationrates}. Filled diamonds mark results from the various sub-regions and the large open diamonds mark the results from the spectra extracted over the full continuum region.}
\label{fig_zetaH}
\end{figure}

From all of the different model runs we construct a set of predicted OH$^+$ and H$_2$O$^+$ column densities for clouds with known $A_V$, $\chi_{\rm UV}$, and $\zeta_{\rm H}$. Using these results we plot contours of constant $\zeta_{\rm H}/n_{50}$ in the parameter space $\log_{10}[N({\rm OH}^+)/N({\rm H})]$ vs. $N({\rm OH}^+)/N({\rm H_2O}^+)$ as shown in Figure \ref{fig_zetaH}, where each panel corresponds to a different value of $\chi_{\rm UV}/n_{50}$ (the use of $n_{50}\equiv n_{\rm H}/[50$~cm$^{-3}$] here clearly highlights the effects of varying density). With $N({\rm OH}^+)$ and $N({\rm H_2O}^+)$ determined from our observations, only $N({\rm H})$ is needed to infer ionization rates from Figure \ref{fig_zetaH}. There are no observations of hydrogen in our target galaxies, so it is necessary to estimate $N({\rm H})$ from observations of another species. The recent observations of CH$^+$ absorption in these galaxies are likely the best tracer of the same gas component giving rise to OH$^+$ and H$_2$O$^+$ absorption \citep{gerin2016}. Surveys of CH$^+$ in the Galaxy have shown that the average relative abundance between CH$^+$ and hydrogen ($X({\rm CH}^+)=N({\rm CH}^+)/N_{\rm H}$, where $N_{\rm H}\equiv N({\rm H})+2N({\rm H}_2)$) in the solar neighborhood is about $X({\rm CH}^+)=7.6\times10^{-9}$ \citep{godard2014}. There is large scatter between individual measurements though, with values between about $2\times10^{-9}$ and $2\times10^{-8}$, meaning any estimate of the hydrogen column density from CH$^+$ will be highly uncertain. Additionally, the molecular hydrogen fraction ($f_{\rm H_2}\equiv 2n({\rm H}_2)/n_{\rm H}$) is required to determine the column density of atomic hydrogen, $N({\rm H})$, given the column density of hydrogen nuclei, $N_{\rm H}$. We adopt the mean value of $f_{\rm H_2}=0.053\pm0.026$ reported by \citet{indriolo2015oxy} in gas containing OH$^+$ and H$_2$O$^+$ for use in calculating $N({\rm H})$ via
\begin{equation}
N({\rm H})=(1-f_{\rm H_{2}})\frac{N({\rm CH}^+)}{X({\rm CH}^+)}.
\label{eq_hcolumn}
\end{equation}
Average column densities of CH$^+$ toward SMM~J2135$-$0102 and SDP~17b (i.e., determined from spectra extracted from the solid angle encompassing all of the continuum emission) are reported as $N({\rm CH}^+)=(8.3\pm1.4)\times10^{14}$~cm$^{-2}$ and $N({\rm CH}^+)=(4.7\pm1.9)\times10^{14}$~cm$^{-2}$, respectively \citep{falgarone2017}, but for the same reasons discussed in Section \ref{section_column} these may be lower limits. With this caveat we estimate $N({\rm H})=(1.0^{+3.6}_{-0.7})\times10^{23}$~cm$^{-2}$ and $N({\rm H})=(5.9^{+25}_{-4.5})\times10^{22}$~cm$^{-2}$ for SMM~J2135$-$0102 and SDP~17b, respectively. Uncertainties are dominated by the scatter in $X({\rm CH}^+)$ as we adopt the limiting values of $2\times10^{-9}$ and $2\times10^{-8}$ in our analysis. Additional uncertainties are introduced if $f_{\rm H_2}$ is larger than the adopted value. The maximum value of $f_{\rm H_2}$ in portions of the model clouds where most of the OH$^+$ and H$_2$O$^+$ reside is about 0.3, so this would cause about a 30\% overestimate in $N({\rm H})$. Note that these values of $N({\rm H})$ exceed the expected total column densities of our thickest model clouds \citep[$N_{\rm H}=1.4\times 10^{22}$~cm$^{-2}$, assuming $N_{\rm H}/A_{V}=1.8\times 10^{21}$~cm$^{-2}$~mag$^{-1}$;][]{predehl1995}, necessitating absorption from multiple clouds along the line of sight.

While we have measured $N({\rm OH}^+)$ and $N({\rm H_2O}^+)$ in each sub-region and have reported their ratio in Table \ref{tbl_measure}, the synthesized beams from observations of each molecule in the same galaxy do not match, as seen in Figure \ref{fig_continuum_images}. As a result, the spatial extent probed by our defined regions is slightly different for OH$^+$ and H$_2$O$^+$, and should be considered a further source of uncertainty in our analysis. The mismatch is more extreme for OH$^+$ and CH$^+$ as the latter only has a single average column density reported for each galaxy. Keeping in mind these caveats, our estimates of $\log_{10}[N({\rm OH}^+)/N({\rm H})]$ and $N({\rm OH}^+)/N({\rm H_2O}^+)$ from observations are plotted in the various panels of Figure \ref{fig_zetaH}. 

\section{Discussion} \label{section_discussion}

The different model grids shown in Figure \ref{fig_zetaH} demonstrate that the inferred cosmic-ray ionization rate is dependent upon the adopted strength of the UV radiation field for $\chi_{\rm UV}/n_{50}\leq10$. It is difficult to determine the strength of the UV radiation field in the gas where the OH$^+$ and H$_2$O$^+$ absorption occurs, but we can attempt to make some rough estimates. \citet{danielson2011} report a UV field over 1000 times\footnote{This study describes the UV flux in terms of the Habing field, $G_0$. At $\chi_{\rm UV}=1$, $G_0=1.7$.} stronger in SMM~J2135$-$0102 than in the Milky Way based on the luminosity ratios of various CO, C, and C$^+$ emission lines. This estimate applies to the dense ($n_{\rm H}\sim10^4$~cm$^{-3}$) gas in close proximity to the starburst regions though, and likely not the gas with $n_{\rm H}\sim50$~cm$^{-3}$ where we see CH$^+$, OH$^+$ and H$_2$O$^+$ absorption. The $\chi_{\rm UV}/n_{50}=1000$ model (panel e in Figure \ref{fig_zetaH}) suggests cosmic-ray ionization rates in SMM~J2135$-$0102 are on the order of $10^{-16}$~s$^{-1}$, several orders of magnitude below the $\zeta_2\sim10^{-13}$--$10^{-11}$~s$^{-1}$ estimate of \citet{danielson2013} for the same gas component were they find the high UV field. This discrepancy suggests that OH$^+$ and H$_2$O$^+$ absorption does not come from gas close to the starburst regions, reaffirming the idea that it instead is located in the same extended halo where CH$^+$ absorption is seen. The fact that the SMM~J2135$-$0102 results occupy the same location in $\log_{10}[N({\rm OH}^+)/N({\rm H})]$ vs. $N({\rm OH}^+)/N({\rm H_2O}^+)$ parameter space as Galactic diffuse clouds \citep{indriolo2015oxy,neufeld2017} also supports this picture. Even if we assume that the UV field is highly attenuated in this extended halo and use the model with $\chi_{\rm UV}/n_{50}=0.1$ (panel a in Figure \ref{fig_zetaH}), ionization rates in SMM~J2135$-$0102 are still only on the order of a few times $10^{-15}$~s$^{-1}$. However, a decrease in the cosmic-ray ionization rate between the starburst regions and an extended halo is in fact expected, since the underlying cosmic-ray flux will decrease with propagation into a much larger volume and due to losses as particles interact with the ambient medium. Assuming the underlying cosmic-ray spectrum does not vary in shape, and that particles propagate away from the star-forming regions isotropically, the cosmic-ray flux and ionization rate at large distances will decrease as $d^{-2}$. The low-energy ($\sim$MeV) particles most efficient at ionization (due to larger interaction cross sections), will be removed from the particle spectrum more quickly due to energy losses \citep{cravens1978,padovani2009} and may be preferentially confined both near the star-forming regions and within the galaxy by magnetic fields \citep[see review by][]{cesarsky1980}, so the underlying cosmic-ray spectrum is expected to change. As a result, the cosmic-ray ionization rate should decrease even faster than $d^{-2}$, and our estimate of $\zeta_{\rm H}\approx10^{-15}$~s$^{-1}$ in the extended gaseous halo around SMM~J2135$-$0102 can be reconciled with the $\zeta_2\sim10^{-13}$--$10^{-11}$~s$^{-1}$ estimate of \citet{danielson2013} in the star-forming region.

There are no previous estimates of the cosmic-ray ionization rate or the strength of the UV radiation field in SDP~17b, and observations of CO and H$_2$O do not resolve the emitting region  \citep{omont2011,omont2013,yang2017}. If we again assume that the OH$^+$ and H$_2$O$^+$ absorption arises in an extended halo where the UV field is weak, we find ionization rates on the order of $10^{-15}$--$10^{-14}$~s$^{-1}$. As for SMM~J2135$-$0102, the cosmic-ray ionization rate in the more compact star-forming regions of SDP~17b where CO and H$_2$O emission have been observed would likely be a few orders of magnitude above our estimates in the extended halo.

Cosmic-ray ionization rates have been inferred in a variety of regions both within and beyond the Milky Way. Diffuse gas in the Galactic ISM has an average ionization rate of about $3\times10^{-16}$~s$^{-1}$ \citep{gerin2010,neufeld2010,indriolo2012,indriolo2015oxy}. Dense gas shows a lower ionization rate of about $3\times10^{-17}$~s$^{-1}$ \citep[e.g.,][]{caselli1998,maret2007}, while gas in the Galactic center shows a higher ionization rate around 10$^{-15}$--10$^{-13}$~s$^{-1}$ \citep[e.g.,][]{yusefzadeh2013,goto2014,lepetit2016}, demonstrating that $\zeta_{\rm H}$ can vary by a few orders of magnitude within a single galaxy. This is not surprising given that the cosmic-ray spectrum at any given location is dependent on proximity to cosmic-ray accelerators, particle propagation effects, and losses via interactions with the ambient medium. Observations of OH$^+$ and H$_2$O$^+$ that probe the nuclear regions of ULIRGs suggest ionization rates of $\zeta_{\rm H}\gtrsim10^{-13}$~s$^{-1}$ \citep{gonzalez-alfonso2013,gonzalez-alfonso2018arxiv}, while OH$^+$ and H$_2$O$^+$ observed in the disks of starburst galaxies give an average value of $\zeta_{\rm H}\approx4\times10^{-16}$~s$^{-1}$ \citep{vandertak2016}. \citet{muller2016} detected OH$^+$ and H$_2$O$^+$ absorption in the disk of the Milky Way-like absorber at $z=0.89$ toward PKS~1830$-$211, and inferred ionization rates in the range $3\times10^{-15}$~s$^{-1}\lesssim \zeta_{\rm H}\lesssim 2\times10^{-14}$~s$^{-1}$. Clearly, there is large variation in cosmic-ray ionization rates inferred in different regions, but the general trend is that $\zeta_{\rm H}$ is larger in regions of more copious star formation. 

The intent of our study was to investigate this relationship in SMGs, which show some of the highest star-formation rates in the Universe. Note that while we have referred to an ionization rate of $10^{-13}$--$10^{-11}$~s$^{-1}$ in SMM~J2135$-$0102 \citep{danielson2013}, that estimate is based on the galaxy's star-formation rate, and so should not be used when investigating the relationship between the two parameters. Although we have not directly determined $\zeta_{\rm H}$ in the star-forming regions of SMGs, the ionization rates of $10^{-16}$--$10^{-14}$~s$^{-1}$ that we infer in the extended gaseous halos around SMM~J2135$-$0102 and SDP~17b are interesting. The fact that the ionization rate is expected to decrease faster than $d^{-2}$ with increasing distance from the site of particle acceleration suggests that cosmic-ray ionization rates in the star-forming regions of SMM~J2135$-$0102 and SDP~17b may be orders of magnitude larger than in the gas probed by our observations. This is an extremely rough estimate, but it is consistent with the trend of increasing $\zeta_{\rm H}$ with increasing SFR.

\section{Summary}

We have used ALMA to search for OH$^+$ and H$_2$O$^+$ absorption in the lensed sub-millimeter galaxies SMM~J2135$-$0102 and SDP~17b at redshift $z\sim2.3$. Both molecules are detected, and their absorption profiles change with position in front of the background continuum sources. Average column densities of OH$^+$ and H$_2$O$^+$ were extracted from multiple regions toward each galaxy, and these have been used in concert with chemical models to infer cosmic-ray ionization rates on the order of 10$^{-16}$--10$^{-14}$~s$^{-1}$ in the extended gaseous halos surrounding these galaxies. Given the changes expected in the cosmic-ray spectrum as particles propagate away from their acceleration sites, ionization rates in the compact star-forming regions of SMM~J2135$-$0102 and SDP~17b are likely orders of magnitude higher. A larger sample of SMGs must be observed to better study the relationship between $\zeta_{\rm H}$ and SFR in this class of galaxies, but our first indications are that ionization rates in SMGs continue to follow the trend with star-formation rate that is observed in the Milky Way and other nearby galaxies.

N.~I. acknowledges the staff at NRAO for help with data reduction, and thanks the anonymous referee for suggestions to improve the paper. M.~G.~W. and D.~A.~N. gratefully acknowledge the support of a grant from NASA's Astrophysical Data Analysis Program (ADAP). This paper makes use of the following ALMA data: ADS/JAO.ALMA\#2016.1.00285.S,  ADS/JAO.ALMA\#2012.1.00175.S and ADS/JAO.ALMA\#2013.1.00569.S. ALMA is a partnership of ESO (representing its member states), NSF (USA) and NINS (Japan), together with NRC (Canada) and NSC and ASIAA (Taiwan) and KASI (Republic of Korea), in cooperation with the Republic of Chile. The Joint ALMA Observatory is operated by ESO, AUI/NRAO and NAOJ. The National Radio Astronomy Observatory is a facility of the National Science Foundation operated under cooperative agreement by Associated Universities, Inc.

\bibliographystyle{aasjournal}
\bibliography{indy_master}


\clearpage
\begin{deluxetable}{cccccc}
\tablecaption{Target Galaxies\label{tbl_targets}}
\tablehead{
\colhead{} & \colhead{} & \colhead{} & \colhead{} & \colhead{$\Sigma_{\rm SFR}$} &  \\
\colhead{Name} & \colhead{IAU Name} & \colhead{$z$} & \colhead{$z$ Ref.} & \colhead{(M$_{\odot}$~yr$^{-1}$~kpc$^{-2}$)} & \colhead{$\Sigma_{\rm SFR}$ Ref.}
}
\startdata
SDP~17b & H-ATLAS J090302.9$-$014127 & 2.3051 & L12 & $52\pm36$ & Y17 \\
Eyelash & SMM J2135$-$0102 & 2.3259 & S10 & $130\pm40$ & D11 
\enddata
\tablecomments{References: L12: \citet{lupu2012};  S10: \citet{swinbank2010}; Y17: \citet{yang2017}; D11: \citet{danielson2011}}
\end{deluxetable}

\begin{deluxetable}{lccccccccc}
\tablewidth{0pt}
\tabletypesize{\scriptsize}
\tablecaption{Transition Properties\label{tbl_transitions}} 
\tablehead{Molecule &  \multicolumn{3}{c}{Transition}   & Rest Frequency & $E_{l}/k$ & $g_u$ & $g_l$ & $A$  \\
& & & & (MHz) & (K) & & & (10$^{-2}$~s$^{-1}$)}
\startdata
  & $N'$--$N''$ & $J'$--$J''$ & $F'$--$F''$ &  &  &  &  &  \\ 
OH$^+$ & 1--0 & 1--1 & 1/2--1/2 & 1032998.5 & 0.0055 & 2 & 2 & 1.41 \\
OH$^+$ & 1--0 & 1--1 & 3/2--1/2 & 1033004.4 & 0.0055 & 4 & 2 & 0.35 \\
OH$^+$ & 1--0 & 1--1 & 1/2--3/2 & 1033112.9 & 0 & 2 & 4 & 0.70 \\
OH$^+$ & 1--0 & 1--1 & 3/2--3/2 & 1033118.6\tablenotemark{a} & 0 & 4 & 4 & 1.76 \\
\hline
\hline
  & $N'_{K'_{a}K'_{c}}$--$N''_{K''_{a}K''_{c}}$ & $J'$--$J''$ & $F'$--$F''$ &  &  &  &  &  \\
$o$-H$_2$O$^+$ 	& $1_{11}$--$0_{00}$ & 1/2--1/2 & 1/2--1/2 & 1139536.6 & 0.0053 & 2 & 2 & 0.37 \\
$o$-H$_2$O$^+$ 	& $1_{11}$--$0_{00}$ & 1/2--1/2 & 3/2--1/2 & 1139555.8 & 0.0053 & 4 & 2 & 1.48 \\
$o$-H$_2$O$^+$ 	& $1_{11}$--$0_{00}$ & 1/2--1/2 & 1/2--3/2 & 1139648.7 & 0 & 2 & 4 & 2.93 \\
$o$-H$_2$O$^+$ 	& $1_{11}$--$0_{00}$ & 1/2--1/2 & 3/2--3/2 & 1139667.9\tablenotemark{a} & 0 & 4 & 4 & 1.83 \\
\enddata
\tablecomments{All data were obtained from the Cologne Database for Molecular Spectroscopy \citep[CDMS;][]{muller2005} on 2017-10-27. OH$^+$ data are from \citet{bekooy1985}.  H$_2$O$^+$ data are from \citet{murtz1998}, with minor frequency shifts due to the findings of \citet{neufeld2010} and \citet{muller2016}.}
\tablenotetext{a}{Indicates the strongest of the hyperfine transitions for a specific $\Delta J$ which was used to set the velocity scale during our analysis.}
\end{deluxetable}
\normalsize

\clearpage
\begin{deluxetable}{lcccccc}
\tablewidth{0pt}
\tabletypesize{\scriptsize}
\tablecaption{Absorption Line Properties and Inferred Parameters\label{tbl_measure}}
\tablehead{ \colhead{Region}  & \colhead{vel. range} & \colhead{OH$^+ \int \tau dv$} & \colhead{H$_2$O$^+ \int \tau dv$} & \colhead{$N({\rm OH}^+)$} & \colhead{$N({\rm H}_{2}{\rm O}^{+})$} & 
\colhead{$N({\rm OH}^+)$/$N({\rm H}_{2}{\rm O}^{+})$} \\
 & \colhead{(km~s$^{-1}$)} & \colhead{(km~s$^{-1}$)} & \colhead{(km~s$^{-1}$)} & \colhead{($10^{14}$~cm$^{-2}$)} & \colhead{($10^{14}$~cm$^{-2}$)} &  } 
\startdata
\multicolumn{7}{c}{\bf SMM J2135$-$0102} \\											
Full	 & 	[$-$220, 610]	 & 	264.6$\pm$3.5	 & 	69.3$\pm$1.8	 & 	12.88$\pm$0.17	 & 	3.86$\pm$0.10	 & 	3.33$\pm$0.10 \\
NW	 & 	[$-$220, 610]	 & 	259.6$\pm$5.1	 & 	69.0$\pm$1.7	 & 	12.63$\pm$0.25	 & 	3.85$\pm$0.09	 & 	3.28$\pm$0.10 \\
NW1	 & 	[$-$220, 280]	 & 	...	 		& 	43.6$\pm$8.4	 & ...		 & 	2.43$\pm$0.47	 & ...	 \\
NW2	 & 	[$-$150, 280]	 & 	231.1$\pm$103.1	 & 	50.5$\pm$6.6	 & 	11.25$\pm$5.02	 & 	2.82$\pm$0.37	 & 	4.00$\pm$1.86 \\
NW3	 & 	[$-$150, 280]	 & 	341.8$\pm$29.3	 & 	62.8$\pm$2.7	 & 	16.64$\pm$1.43	 & 	3.50$\pm$0.15	 & 	4.75$\pm$0.46 \\
NW4	 & 	[$-$220, 340]	 & 	373.9$\pm$20.1	 & 	98.2$\pm$2.8	 & 	18.20$\pm$0.98	 & 	5.47$\pm$0.16	 & 	3.32$\pm$0.20 \\
NW5	 & 	[$-$150, 450]	 & 	361.5$\pm$11.5	 & 	103.9$\pm$1.6	 & 	17.59$\pm$0.56	 & 	5.79$\pm$0.09	 & 	3.04$\pm$0.11 \\
NW6	 & 	[$-$130, 610]	 & 	398.4$\pm$10.1	 & 	86.1$\pm$2.6	 & 	19.39$\pm$0.49	 & 	4.80$\pm$0.14	 & 	4.04$\pm$0.16 \\
NW7	 & 	[$-$130, 610]	 & 	255.1$\pm$20.6	 & 	67.4$\pm$3.0	 & 	12.42$\pm$1.00	 & 	3.76$\pm$0.17	 & 	3.30$\pm$0.30 \\
NW8	 & 	...	 & 	...			 & ...		 & ...		 & ...		 & ...	 \\
SE	 & 	[$-$220, 610]	 & 	247.2$\pm$4.4	 & 	66.4$\pm$3.0	 & 	12.03$\pm$0.21	 & 	3.70$\pm$0.17	 & 	3.25$\pm$0.16 \\
SE1	 & 	[$-$220, 280]	 & 	...			 & 	33.8$\pm$8.7	 & ...		 & 	1.88$\pm$0.48	 & ...	 \\
SE2	 & 	[$-$150, 280]	 & 	166.4$\pm$46.1	 & 	21.1$\pm$6.7	 & 	8.10$\pm$2.24	 & 	1.18$\pm$0.37	 & 	6.89$\pm$2.90 \\
SE3	 & 	[$-$150, 280]	 & 	234.8$\pm$25.4	 & 	46.4$\pm$5.4	 & 	11.43$\pm$1.24	 & 	2.59$\pm$0.30	 & 	4.42$\pm$0.70 \\
SE4	 & 	[$-$220, 340]	 & 	420.3$\pm$26.3	 & 	88.3$\pm$3.1	 & 	20.46$\pm$1.28	 & 	4.92$\pm$0.17	 & 	4.16$\pm$0.30 \\
SE5	 & 	[$-$150, 450]	 & 	393.7$\pm$27.2	 & 	89.1$\pm$4.0	 & 	19.16$\pm$1.32	 & 	4.97$\pm$0.22	 & 	3.86$\pm$0.32 \\
SE6	 & 	[$-$130, 610]	 & 	336.3$\pm$13.7	 & 	70.8$\pm$4.6	 & 	16.37$\pm$0.67	 & 	3.95$\pm$0.26	 & 	4.15$\pm$0.32 \\
SE7	 & 	[$-$130, 610]	 & 	164.1$\pm$14.9	 & 	63.8$\pm$7.3	 & 	7.99$\pm$0.73	 & 	3.56$\pm$0.41	 & 	2.25$\pm$0.33 \\
SE8	 & 	...	 & 	...			 & ...		 & ...		 & ...		 & ...	 \\
\hline
\multicolumn{7}{c}{\bf SDP~17b} \\	
Full	 & 	[$-$755, 235]	 & 	339.6$\pm$3.8	 & 	22.8$\pm$2.6	 & 	16.53$\pm$0.18	 & 	1.27$\pm$0.14	 & 	13.00$\pm$1.49 \\
North & 	[$-$755, 235]	 & 	275.2$\pm$7.5	 & 	33.0$\pm$4.7	 & 	13.39$\pm$0.37	 & 	1.84$\pm$0.26	 & 	7.28$\pm$1.06 \\
East	 & 	[$-$755, 235]	 & 	323.0$\pm$10.6	 & 	23.6$\pm$2.6	 & 	15.72$\pm$0.52	 & 	1.32$\pm$0.14	 & 	11.95$\pm$1.37 \\
South & 	[$-$755, 235]	 & 	626.6$\pm$16.4	 & 	57.0$\pm$4.6	 & 	30.50$\pm$0.80	 & 	3.18$\pm$0.26	 & 	9.60$\pm$0.81 \\
West	 & 	[$-$755, 235]	 & 	374.4$\pm$4.6	 & 	30.0$\pm$3.6	 & 	18.22$\pm$0.22	 & 	1.67$\pm$0.20	 & 	10.90$\pm$1.31 \\
\enddata
\tablecomments{Integrated optical depths are determined over the tabulated velocity ranges from the spectra in Figures \ref{fig_eyelash_ohp_spectra_v2} through \ref{fig_sdp17b_spectra}. Column densities are calculated from equation (\ref{eq_column}) as described in Section \ref{section_column}. Measured quantities in corresponding mirrored regions toward SMM~J2135$-$0102 are in agreement within 3$\sigma$ uncertainties.}
\end{deluxetable}
\normalsize


\clearpage

\clearpage

\clearpage

\clearpage

\end{document}